\documentclass[prd,amsmath,amsfonts,preprint,11pt,superscriptaddress,onecolumn]{revtex4}

\usepackage[dvips]{color}
\usepackage{amsmath}
\usepackage{amsfonts}
\usepackage{amssymb}

\usepackage{graphics}
\usepackage{amsmath}
\usepackage{amsfonts}
\usepackage{amssymb}
\usepackage{graphicx}
\usepackage{epsfig}

\def\ba{\begin{eqnarray}}
\def\ea{\end{eqnarray}}
\def\be{\begin{equation}}
\def\ee{\end{equation}}

\def\d{\mathrm{d}}

\def\mn{_{\mu \nu}}
\def\mupn{^\mu_{\, \nu}}
\def\({\left(}
\def\){\right)}

\def\rt{\tilde{r}_\pm}
\def\bd{\boxdot}
\def\rrp{\tilde{r}_+}
\def\rrm{\tilde{r}_-}

\newcommand{\ind}[2]{\!\!\!\phantom{#1}^{(#2)}\!#1}

\begin{document}

\title{Gravitational Waves in a Codimension Two Braneworld}

\author{Claudia de Rham}
\email{derham@hep.physics.mcgill.ca}
\affiliation{Department of Applied Mathematics and Theoretical Physics,
University of Cambridge, Wilberforce Road, Cambridge CB3 0WA, England.}
\affiliation{Ernest Rutherford Physics Building, McGill University,
  3600 rue University, Montreal, QC H3A 2T8, Canada.}
\author{Andrew J. Tolley}
\email{atolley@princeton.edu}
\affiliation{Joseph Henry Laboratories, Princeton University, Princeton NJ,
08544, USA.}

\date{\today}

\begin{abstract}
We consider the propagation of gravitational waves in six dimensions
induced by sources living on 3-branes in the context
of a recent exact solution \cite{Mukohyama:2005yw}.
The brane geometries are de Sitter and the bulk is a warped geometry
supported by a positive cosmological
constant as well as a 2-form flux. We show that at low energies
ordinary gravity is reproduced, and explicitly compute the leading
corrections from six dimensional effects. After regulating the brane
we find a logarithmic dependence on the cutoff scale of brane physics
even for modes whose frequency is much less than this energy scale. We
discuss the possibility that this dependence can be renormalized into
bulk or brane counterterms in line with effective field theory
expectations. We
discuss the inclusion of Gauss-Bonnet terms that have been used
elsewhere to regulate codimension two branes. We find that such terms
do not regulate codimension two branes for compact extra dimensions.
\end{abstract}
\maketitle
%
\section{Introduction}
In this paper we discuss a simple six dimensional model, with two
compact extra dimensions with matter living on two conical 3-branes. This model
may well approximate an inflationary phase, or late-time
dark energy dominated phase in scenarios with two large extra dimensions as
proposed in the original scenario of Arkani-Hamed, Dimopolous and Dvali (ADD)
\cite{Arkani-Hamed:1998rs,Antoniadis:1998ig}.
Our concern shall be how the matter on the brane backreacts on the
bulk, and in particular the resulting gravitational waves that
are generated. We find that at low energies, at leading order ordinary
gravity is reproduced with the anticipated effective Newton
constant. In addition there are a series of corrections suppressed by
powers of $R_c^2 E^2$ where $R_c$ is the size of the extra
dimensions. The magnitude of the leading order correction is sensitive to how we
regulate the brane even at energy scales far below $E \ll 1/l$ where
$l$ is the width of the brane. This suggests an apparent model
dependence in dealing with codimension two branes. We discuss the
possibility that this dependence can be renormalized, i.e. absorbed
into brane and bulk counterterms. Recently similar ideas have been
proposed as being useful for understanding black hole physics
\cite{Goldberger:2004jt}. 

The proposal of ADD resolves the hierarchy problem between the Planck
and electroweak scales by having the fundamental Planck mass of order the
electroweak scale, and using the dilution effect of gravity in extra
dimensions to give rise to an effectively weak four-dimensional gravity. In the
context of string theory this typically introduces a new hierarchy
between the size of several large extra dimensions and additional
small extra dimensions and/or the fundamental Planck scale, something
which may or may not be natural depending on the details of the moduli
stabilization. In the supersymmetric large extra dimension scenario
(SLED) it has been suggested that the same features of the large extra
dimensions combined with the calming effects of SUSY can also be used
to alleviate the cosmological constant hierarchy problem
\cite{Aghababaie:2003wz,Rubakov:1983bz}, (see \cite{Burgess:2005wu} for
a recent review). A detailed understanding of whether this is true
requires amongst other things an understanding of how changes in the brane tensions, of the type
that may arise in phase transitions on the brane, influence the bulk
dynamics. It has proven technically challenging \cite{Rubakov:1983bz}
to answer this question for reasons that shall become apparent in the
following. Here we shall focus on the simpler question of how the bulk
dynamics is influenced by small matter perturbations on the brane. 

Although great progress has been made in recent years in
understanding the dynamics of codimension one branes, higher
codimension branes remain something of an enigma. This arises
because at the level of GR, distributional sources of arbitrary
codimension are typically singular \cite{Geroch:1987qn,Corradini:2002ta}. As a
result one needs to deal with `thick' branes such as those described
by regular defects from field theory models
\cite{Vilenkin:1981zs,Gherghetta:2000qi,Bazeia:2003qt}. An alternative approach
is to modify Einstein gravity by the addition of higher derivative
curvature terms which can potentially allow distributional sources to
be well defined. In the case of codimension two branes one such
approach that has been taken is to use Gauss-Bonnet terms
\cite{Bostock:2003cv,Corradini:2001su,Kofinas:2005py,Arkani-Hamed:1999hk}.
One of the main motivations for the use of these terms comes
from string theory where they generically appear as a leading order
quantum correction to gravity
and guarantee a ghost-free action
\cite{Zwiebach:1985uq}.
However, as
will become apparent later this method of regularizing the branes with Gauss-Bonnet terms can only be applied for
codimension two branes in a noncompact extra dimension, and is
inconsistent for the more familiar case of compact extra
dimensions. Some other inconsistencies have been suggested in the
case of an axially symmetric bulk, where an isotropic braneworld
cosmological ansatz seems to be incompatible with the  model
\cite{Kofinas:2005py}.

In string theory we are
interested in the dynamics of D-branes of arbitrary dimension, and
here we find that at the level of the supergravity, the geometry
describing all but the D3-branes are singular. In string theory we
are to understand that string $\alpha'$ or $g_s$ corrections will
`regulate' the singularity. However, at first sight there seems to
be a contradiction with expectations from low energy effective field
theory (EFT), since this suggests that we need to understand the
string scale physics that regulates the brane in order to make
predictions about low energy dynamics of relevance to cosmology.
It is precisely this aspect that we would like to explore in the
following by asking the simpler question, how does the production of
gravitational waves induced by sources living on the brane depend on
the detailed physics that resolves the brane?

We begin in section \ref{sectionII} by discussing approaches to
regulating the brane geometries. Then in section \ref{sectionIII} we
introduce the background solution and calculate the tensor
perturbations via a derivative expansion. This enables us to obtain
order by order the modified equations for the gravitational waves on
the brane in section \ref{sectionIV}. We discuss the presence of the logarithmic divergences that
arise in the thin brane limit and whether these can be canceled by
counterterms on the brane or in the bulk. After reviewing in detail the
Kaluza-Klein limit of the general solution in section
\ref{sectionV},
we compute in section \ref{GB}, the effects of
adding a Gauss-Bonnet term in the
bulk and show that this cannot be used to regulate the divergences near the brane.
Finally in section \ref{conclusion} we conclude.

\section{Dealing with Codimension Two Branes}

\label{sectionII}

In this section we discuss our approach to dealing with codimension two
branes. We follow closely the approach of ref.
\cite{Navarro:2004di}.

\subsection{Thick Branes}

\label{boundaryconditions2}

\begin{figure}[h]
 \begin{center}
 \includegraphics[width=15cm]{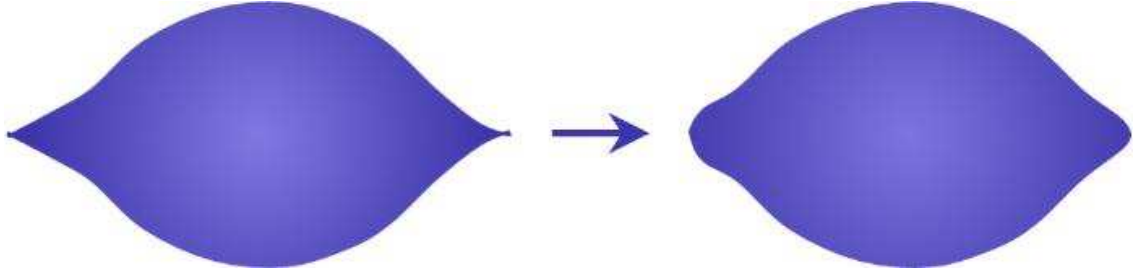}\\
 \caption{Six dimensional compactified spacetime with regularized
 singularities.}
 \label{figure regular}
 \end{center}
 \end{figure}

One approach to dealing with codimension two objects is to regulate
them by replacing them with a smooth stress energy source, i.e. a
`thick brane' \cite{Corradini:2002ta,Gregory:1995qh}. For
instance one model of this would be as a
codimension two topological defect arising for example from a six
dimensional Abelian-Higgs model \cite{Vilenkin:1981zs}. Figure \ref{figure regular}
illustrates what the resulting geometry of the two extra dimensions would
look like.
The conical deficits at the poles are replaced with smooth geometries.

Although the full behaviour of the solution in the polar regions
may be quite complicated,
provided the width $l$ of the brane is much less than the scales
of physical interest we
anticipate that the bulk physics will only be sensitive to integrated
effects over the brane, a perspective espoused in \cite{Navarro:2004di}.
As in many works on this topic we shall restrict ourselves to
axisymmetric solutions which is a consistent (if not entirely
desirable) truncation. Given a
six dimensional metric of the form
\be
\d s^2=N(r)^2 \d r^2+\mathcal{L}(x,r)^2\d\varphi^2 + g_{\mu\nu}(x,r)\d
x^{\mu}\d x^{\nu},\label{general 6d metric}
\ee
with branes located at $r_{-}$ and $r_+>r_-$, we define the four dimensional
stress energy of the brane by
\be
{}^{(4)}T^{\mu}_{\nu}{}^{(\pm)}=
\frac{1}{\sqrt{-g(r=\rt)}}\int_{r_{\pm}}^{\rt}
\d r \d\varphi \, N(r) \mathcal{L} (r,x) \sqrt{-g} \, {}^{(6)}T^{\mu}_{\nu},
\ee
with $\tilde{r}_{\pm}=r_{\pm} \mp \epsilon$. Here
${}^{(6)}T^{\mu}_{\nu}$ is the regular stress energy source describing
the thick brane
which we assume vanishes outside the region $|r-r_{\pm}| \le
\epsilon$. Here $\epsilon$ is related to the width of the brane by 
 $l=|\int_{r_{\pm}}^{\rt} N(r) dr|$.
The requirement that the
 metric is smooth as $r \rightarrow r_{\pm}$ implies
\ba
\mathcal{L} (x,r_{\pm})&=&0,  \\
N(r_{\pm})^{-1} \partial_r \mathcal{L} (x,r_{\pm})&=&\mp 1 , \\
 N(r_{\pm})^{-1}\partial_r g_{\mu\nu}(x,r_{\pm})&=&0.
\ea
The key observation of \cite{Navarro:2004di} is that by integrating
the Einstein equations over the brane thickness, one may infer an
effective matching rule for the extrinsic curvature on the surface
$r=r_{\pm}\mp \epsilon$ in terms of the integrated brane stress
energies. This matching rules play the role of the Israel junction
conditions for codimension one branes.

In particular if one looks at the ${}^{(6)}R^{\mu}_{\nu}$ equation we
have \footnote{Our convention is such that the Einstein action is
  $S=\frac{1}{2\kappa}\int \d^6x\sqrt{-g} \ \, {}^{(6)\!}R$ and the metric is `mostly plus'.}
\be
{}^{(6)}R^{\mu}_{\nu}={}^{(4)}R^{\mu}_{\nu}-\mathcal{L}
^{-1}\nabla^{\mu}\nabla_{\mu}\mathcal{L} -\frac{1}{N\sqrt{-g}}\,
\partial_r
\(\sqrt{-g}\mathcal{L}  K^{\mu}_{\nu}\)=\kappa
\({}^{(6)}T^{\mu}_{\nu}-\frac{1}{4}\delta^{\mu}_{\nu}{}^{(6)}T^{M}_{M}\).
\ee
Here $K^{\mu}_{\nu}$ is the extrinsic curvature defined by $K^A_B=\frac{1}{2N}\, g^{AC}\partial_r g_{CB}$.
The dominant contribution is expected to come from the second order
derivatives, and so on integrating around the brane we find
\be
2\pi \mathcal{L} (\rt) K^{\mu}_{\nu}|_{r=\rt} = \pm \kappa
\({}^{(4)}T^{\mu}_{\nu}{}^{(\pm)}-\frac{1}{4}\delta^{\mu}_{\nu}{}^{(4)}T^{M}_{M}{}^{(\pm)}\)
+\mathcal{O}(\epsilon).
\ee
Similarly we have
\be
2\pi \mathcal{L} (\rt) K^{\varphi}_{\varphi}|_{r=\rt} \pm \pi
\frac{\sqrt{-g}|_{r=r_{\pm}}}{{\sqrt{-g}|_{r=\rt}}}
= \pm \kappa
\({}^{(4)}T^{\varphi}_{\varphi}{}^{(\pm)}-\frac{1}{4}{}^{(4)}T^{M}_{M}{}^{(\pm)}\)+\mathcal{O}(\epsilon).
\ee
Note that the $T^{\mu}_{\varphi}$ terms all vanish due to the initial assumption of the form of the metric.

\subsection{Regularizing as a Codimension One Brane}
\begin{figure}[h]
\begin{center}
\includegraphics[width=7cm]{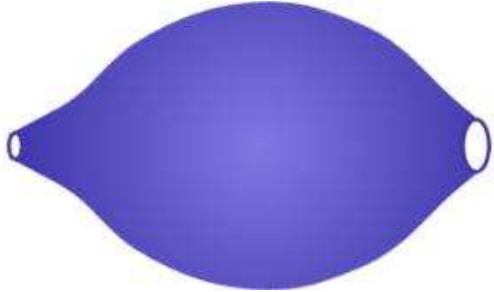}\\
\caption{Six dimensional compactified spacetime with codimension-one orbifold
 planes cutting off the singularity}
\label{figure regbrane}
\end{center}
\end{figure}

%
%
The previous matching conditions amount to a statement about the
extrinsic curvature defined on a codimension one surface in terms of
the brane stress energy which is precisely the same physical
information as if we really had a codimension one brane, localized for
example at an orbifold fixed point as in the Randall-Sundrum model as
shown in figure \ref{figure regbrane}.

Consequently we may regulate the codimension two brane as a codimension
one brane located at an orbifold fixed point, where the Israel
matching conditions would give
\ba
K^{\mu}_{\nu}|_{r=\rt} = \pm \frac{1}{2} \kappa
\({}^{(5)}T^{\mu}_{\nu}{}^{(\pm)}-\frac{1}{4}\delta^{\mu}_{\nu}{}^{(5)}T^{M}_{M}{}^{(\pm)}\)+\mathcal{O}(\epsilon)
\\
K^{\varphi}_{\varphi}|_{r=\rt} = \pm
\frac{1}{2} \kappa
\({}^{(5)}T^{\varphi}_{\varphi}{}^{(\pm)}-\frac{1}{4}{}^{(5)}T^{M}_{M}{}^{(\pm)}\)
+\mathcal{O}(\epsilon).
\ea
Note that the factor of $1/2$ arises because of the doubling of the extrinsic curvature at the fixed point.
The five dimensional stress energy is related to the four dimensional stress energy by
\ba
{}^{(5)}T^{\mu}_{\nu} {}^{(\pm)}&=& \frac{1}{\pi \mathcal{L}(x,\rt)
}
\({}^{(4)}T^{\mu}_{\nu}{}^{(\pm)}+\delta^{\mu}_{\nu}\({}^{(4)}T^{r}_{r}{}^{(\pm)}+\frac{\pi}{\kappa}
\frac{\sqrt{-g}|_{r_{\pm}}}
{\sqrt{-g}|_{\rt}}\)\)\\
{}^{(5)}T^{\varphi}_{\varphi} {}^{(\pm)}&=& \frac{1}{\pi
\mathcal{L}(x, \rt) }
\({}^{(4)}T^{\varphi}_{\varphi}{}^{(\pm)}+{}^{(4)}T^{r}_{r}{}^{(\pm)}\).
\ea

\subsection{Tensor perturbations}

One of the problems with the above matching rules is that they only
contain partial information without a specification of ${}^{(4)}T^r_r$
and ${}^{(4)}T^{\varphi}_{\varphi}$. In the following we shall be
interested in tensor perturbations about a fixed cosmological
background (gravitational waves), where tensors are defined with
respect to four dimensional observers on the brane. In this case we have
$\delta T^r_r=\delta T^{\varphi}_{\varphi}=0$ and so
\ba
\delta {}^{(5)}T^{\mu}_{\nu}{}^{(\pm)} &=& \frac{1}{\pi \mathcal{L} (x, \rt)} \ \delta
       {}^{(4)}T^{\mu}_{\nu}{}^{(\pm)} \label{T5 T4}\\
\delta {}^{(5)}T^{\varphi}_{\varphi}{}^{(\pm)} &=& 0.
\ea
It is this simplification that will allow us to determine the bulk solution without further specification
of the physics of the brane other than through the cutoff $\epsilon$.

\section{Six Dimensional Solutions}

\label{sectionIII}

We use the index conventions that Greek
indices are four dimensional, labeling the transverse $x^\mu$
directions, while small Roman indices are five dimensional, labeling the $x^\mu$ and
$\varphi$ coordinates. The full six dimensional coordinates are
represented by capital Roman indices: $A=0,\cdots,5$. Our starting point is the action
\be
S=\int \d^6 x \sqrt{-g} \frac{1}{2 \kappa} \({}^{(6)}R-2 \Lambda -\frac{1}{2} F_{AB}F^{AB} \),
\ee
for gravity and a bulk form field $F_{AB}=\partial_A A_B-\partial_B
A_A$, ($A_B$ being a $U(1)$ gauge field) and cosmological constant
$\Lambda$. For other six dimensional solutions see
\cite{Kanti:2001vb,Gogberashvili:2003ys} and for cosmology on
codimension two branes, see for instance \cite{Hayakawa:2003qm,Cline:2003ak}.

\subsection{Background solution}
We consider a special case of (\ref{general 6d metric}), where fluxes
are present in the six dimensional bulk and $\mathcal{L}(r,x)^2=L^2
f(r)$, leading to the compactification of \cite{Mukohyama:2005yw},
with a metric of the
form
\ba
\d s^2=g_{AB}\d x^A \d x^B
&=&f^{-1}\d r^2+L^2 f \d \varphi ^2+g\mn \d x^\mu\d x^\nu \label{metric} \\
g\mn \d x^\mu\d x^\nu&=&H^2 r^2 q\mn \d x^\mu\d x^\nu,
\ea
where $q\mn$ is the four-dimensional metric.
For the background, $q\mn=\gamma\mn=a^2(\tau)\eta\mn$, $a(\tau)=\(-H \tau\)^{-1}$
and $f=1
-\frac{\Lambda}{10}r^2-\frac{\mu}{r^3}-\frac{b^2}{12 r^6}$, where $\mu,
b$ are some constants and $\Lambda$ is the  six dimensional
cosmological constant. The gauge field is of the form $A_M \d
x^M=\frac{b}{3r^3}\, L \d \varphi$. $H$ is an arbitrary reference
scale that we have included to keep dimension.  $L$ is included so
that we can normalize $\varphi \in [0,2\pi]$. The proper size of the
$\varphi$ direction is then $2\pi \sqrt{f} L$.

The properties of this solution are discussed more fully in
\cite{Mukohyama:2005yw}, the main point is that provided
$\Lambda>0$, one can find solutions for which $f$ vanishes linearly
at two positive points on the real axis $r=r_{\pm}$, and for
arbitrary values of the parameters these points will be locally Minkowski with a conical
deficit angle signifying the positions of two conical 3-branes. To make this clear
one may define $\rho_{\pm}=\sqrt{\frac{\mp
2(r-r_{\pm})}{\sigma_{\pm}}}$ and $\sigma_{\pm}=\mp \frac{1}{2}
f'(r_{\pm})$, for which the near brane metric becomes
\be
\d s^2 \approx \d\rho_{\pm}^2+L^2
\sigma_{\pm}^2 \rho_{\pm}^2 \d\varphi^2+H^2 r_{\pm}^2 q_{\mu\nu}\d x^{\mu}\d x^{\nu}.
\ee
The conical deficit angle is defined through
$2\pi(1-\delta_{\pm})=2\pi L\sigma_{\pm}$ which corresponds to two
branes of tension
\be
T_{\pm}=\kappa^{-1} (1-L\sigma_{\pm}),\label{tension}
\ee
where $\kappa$ is the six-dimensional Newtonian constant.
For fixed bulk cosmological constant $\Lambda$, we are free to
choose $\mu$, $b$ and $L$ which gives us more than enough freedom to
fit two branes of arbitrary tension. In fact we may set the flux to
zero, i.e. $b=0$ and provided $\Lambda>0$ we can still find
solutions with arbitrary brane tension by adjusting $L$ and $\mu$.
The geometry of each brane
is de Sitter with Hubble constants $H_{\pm}=1/r_{\pm}$ which is only
indirectly related to the brane tensions and the bulk flux and
cosmological constant.

\subsubsection*{Boundary conditions}
Following the prescription of section \ref{boundaryconditions2}, we
now consider two codimension one branes located at $r=\rt$. The
normal vector to the branes is $N_A \d x^A=f^{-1/2} \d r$ and the
extrinsic curvature is
\ba
K\mupn = \frac{\sqrt{f}}{r}\, \delta \mupn \hspace{15pt}
K^\varphi_\varphi =\frac{f'}{2\sqrt{f}} \hspace{15pt}
K^\varphi_\mu = 0 \hspace{15pt} K_\varphi^\mu=0.
\ea
On the branes, the following junction condition should be satisfied:
\ba
\Delta \left[K \delta ^a_b-K^a_b\right]_{\rt}=\kappa \, {}^{(5)}T^{a
(\pm)}_{b},
\ea
where $^{(5)}T^{a (\pm)}_b$ is the stress-energy tensor for the gauge and matter
fields introduced on the codimension one branes. The extrinsic curvature on the brane
should hence satisfy
\ba
K^a_b(\rt)=\pm\frac{\kappa}{2} \!\!\phantom{T}^{(5)}\tilde{T}^{a (\pm)}_b
=\pm\frac{\kappa}{2} \({}^{(5)}T^{a
(\pm)}_b-\frac 1 4 {}^{(5)}T^{c (\pm)}_c
\delta^a_b \).
\ea
For the background, we should therefore have:
\ba
{}^{(5)}\tilde{T}^{\mu (\pm)}_\nu&=& \pm \frac{2}{\kappa \,
\rt}\sqrt{f(\rt)}\delta^\mu_\nu\\
{}^{(5)}\tilde{T}^{(\pm)}_{\varphi
\varphi}&=&\pm\frac{L^2}{\kappa}\sqrt{f(\rt)}.
\ea
%
%
\subsection{Tensor perturbations}

We now study the cosmological perturbations around this background solution sourced
by matter on the branes.
To start with, we consider only four-dimensional tensor
perturbations to the background solution. As elsewhere we only
consider axisymmetric perturbations which is a consistent truncation.
We therefore have $q\mn=\gamma\mn+h\mn(x^\mu,r)$
with $\gamma ^{\mu\nu} h\mn=0$ and $\gamma^{\mu \alpha}h_{\alpha \nu ;\,
\mu}=0$.

Working in this gauge, the only non-trivial contribution to the
Ricci tensor is
\ba
^{(6)}R\mn\left[g_{AB}\right]={}^{(4)}R\mn\left[q\mn\right]-H^2\(3f+r
f'\)q\mn
-2H^2r f h_{\mu \nu ,\, r}-\frac{1}{2} r^2 H^2 \left(f h_{\mu \nu ,\, r}\right)_{,r},
\ea
the perturbed Einstein equation is therefore $\delta ^{(6)}\!\!R\mupn=0$
\ba
\partial_r \(f \, \partial_r h^{\mu}_\nu \)
+\frac{4}{r} f\, \partial_r h^{\mu}_\nu= \frac{2}{H^2 r^2}\, \delta
\bar{R}^\mu_\nu=-\frac{1}{H^2 r^2} \boxdot h^\mu_\nu . \label{deltaR=0}
\ea
The operator $\boxdot$ may be expressed as
\ba
\boxdot h\mupn=\Big[\Box-2H^2 \Big] \, h\mupn,
\ea
$\Box$ being the four-dimensional Laplacian:
$\Box=\gamma^{\alpha \beta}{D}_\alpha {D}_\beta$.

\subsubsection*{Boundary conditions}

At the perturbed level, we consider some matter with stress-energy $\delta
\ind{T}{5}^{a(\pm)}_b=g^{ac}(\rt)\, \delta \ind{T}{5}^{(\pm)}_{cb}$ on each brane.
Since we concentrate for now on four-dimensional tensor
perturbations, we impose $\delta \ind{T}{5}^{\varphi (\pm)}_b=0$,
$\delta \ind{T}{5}^{(\pm)}=\delta
\ind{T}{5}^{a(\pm)}_a=\delta \ind{T}{5}^{\mu(\pm)}_\mu=0$ and $\delta
\ind{T}{5}^{\mu(\pm)}_{\nu ; \mu}=0$.

From the junction conditions, we should have: $\delta
K^a_b(\rt)=\pm\frac{\kappa}{2}\delta
\ind{T}{5}^{a(\pm)}_b$ with
\ba
\delta K\mupn &=& \frac{\sqrt{f}}{2}\, \partial_r h\mupn \\
\delta K^\varphi_\varphi=0, \  \ \
\delta K^\varphi_\mu&=&0, \  \ \ \delta K_\varphi^\mu=0.
\ea
This relation implies:
\ba
\left. \frac{\sqrt{f}}{2}h^\mu_{\nu,r} \right|_{\rt}=
\pm \frac{\kappa}{2}\delta \ind{T}{5}^{\mu (\pm)}_\nu, \label{junction}
\ea
we recall that $h^\mu_\nu=\gamma^{\mu\alpha} h_{\alpha \nu}$,
whereas
$\delta \ind{T}{5}^{a(\pm)}_b=g^{ac}(\rt)\, \delta
\ind{T}{5}^{(\pm)}_{cb}$, hence
$\delta \ind{T}{5}^{\mu
(\pm)}_\nu=\frac{1}{H^2\rt^2}\gamma^{\mu\alpha}\delta
\ind{T}{5}_{\alpha \nu}$. From now on, we use the notation
${}^{(5)}\!\tau^{\mu(\pm)}_\nu=\gamma^{\mu\alpha}\delta
\ind{T}{5}_{\alpha
\nu}$, and hence the junction condition is: $\partial_r h^\mu_{\nu}(\rt)=
\pm \frac{\kappa}{H^2 \rt^2 \sqrt{f(\rt)}}\, \ind{\tau}{5}^{\mu
  (\pm)}_{\, \nu}$.
Furthermore, we may recall from
the relation (\ref{T5 T4}) between the five and
the four-dimensional stress-energy tensor that
$\ind{\tau}{5}^{\mu(\pm)}_\nu=\frac{1}{
\pi L \sqrt{f(\tilde{r}_\pm)}}\ind{\tau}{4}^{\mu(\pm)}_\nu$,
we hence have the boundary condition in terms of the
four-dimensional stress-energy tensor:
\ba
\partial_r h^\mu_{\nu}(\rt)=
\pm \frac{\kappa}{\pi L H^2 \rt^2 f(\rt)}\, \ind{\tau}{4}^{\mu
  (\pm)}_{\, \nu}.
\ea
%
%
\section{Low-energy expansion}
\label{sectionIV}
At low-energies, we may consider that the contribution from
the four-dimensional derivatives to be small in comparison to the $r$
derivatives. We can hence consider an expansion in the operator
$\boxdot$.
The way of solving the Einstein equation
(\ref{deltaR=0}), will be similar to the RS case.
For that, we will express the solution of (\ref{deltaR=0}) as an expansion
in $\bd$ and in what follows we omit any indices to lighten the
notation and $h$
will designate $h\mupn$, and similarly for
$\ind{\tau}{4}^{(\pm)}$:
\ba
h(r,x^\mu)=\sum_{n\ge 0} \(\frac{\bd}{H^2}\)^{n-1}\ h_n(r,x^\mu),
\ea
where we consider each $h_n$ to be of order zero in
$\boxdot/H^2$. So each $h_n$ have a similar weight in the
expansion, but they are weighted by a factor $\(\bd/H^2\)^{n-1}$ which makes
their effective contribution smaller and smaller in the low-energy
regime. In particular we will consider the sum to be dominated by
the zero mode $h_0(r,x^\mu)$.

We can now solve the modified Einstein equation for each
$h_n(r,x), \ n\ge0$:
\ba
\partial_r g_n(r,x)+\frac{4}{r}\,
g_n(r,x)=-\frac{1}{r^2}h_{n-1}(r,x),\label{modes
eq}
\ea
with the notation $h_{-1}(r,x)=0$ and
$g_n = f \partial_r h_n$.

Each mode should satisfy as well the junction conditions. Using the
constraint (\ref{junction}), we therefore have the boundary
conditions for each mode:
\ba
\partial_r
h_n(\rt,x^\mu)&=&0\hspace{20pt}\forall\hspace{5pt} n\ge0,
n\ne 1 \label{junc n}\\
\partial_r h_1(\rt,x^\mu)&=&
\pm \frac{\kappa}{ \pi L H^2\rt^2 f(\rt)}\, \ind{\tau}{4}^{(\pm)}.
\label{junc n=1}
\ea
At this point we should stress that in dealing with functions of
$\boxdot$ (or $\Box$) we always have to be careful that there are
implicitly homogeneous solutions of the equations. For instance in
dealing with an equation of the form $[\Box+a \Box^2]h=T$ we shall
write its solution in the form $h=\Box^{-1}(1-a \Box
+O(\Box^2))T$. Whilst this is a particular solution, we should also
include the homogeneous solution satisfying $[\Box+a \Box^2]h=0$. We
shall take it as read in what follows that these homogeneous solutions
should be included, and hence concentrate on the particular
solution. We may point out that this method would give rise to
precisely the same result as obtained in \cite{Garriga:1999yh} if it
was applied to a codimension one brane.
%
%
\subsection{Zeroth order}
The Einstein equation for the zero mode
(\ref{modes eq}) can be easily solved (in particular, it does not
depend on $f$), and the solution is simply
\ba
\partial_r \, h_0(r,x^\mu)
=\frac{D_0(x^\mu)}{r^4 f}\label{h0'}.
\ea
The constant $D_0$ may be fixed
using the junction condition (\ref{junc n}) which fixes $D_0=0$,
and so:
\ba
h_0(r,x^\mu)=C_0(x^\mu).
\label{h0}
\ea
As we shall see in what follows, $C_0$ will be fixed by the
junction condition for the first mode since the zero mode acts as a source term for the first mode.
%
%
\subsection{First order}
The first order mode can be found by solving
\ba
\partial_r \, g_1+\frac{4}{r} \,
g_1=-\frac{1}{r^2}h_0\,
=-\frac{D_1(x^\mu)}{r^2},
\ea
the solution of this equation satisfies
\ba
\partial_r \, h_1=
\frac{g_1}{f}=
-\frac{C_0(x^\mu)}{3 f r}
+\frac{D_1(x^\mu)}{f r^4}.
\label{h1'}
\ea
We work for now in the region $r_- <\tilde{r}_-
<r<\tilde{r}_+<r_+$, and $^{(1)}\!h(r,x^\mu)$ is hence
regular.
The constants $C_0(x^\mu)$ and  $D_1(x^\mu)$
should be fixed from the
junction conditions as follows. The junction condition for the first
mode is given in (\ref{junc n=1}).
We therefore have:
\ba
C_0(x^\mu)&=&\frac{3 \kappa}{\pi L H^2 \(\tilde{r}_-^3-\tilde{r}_+^3 \)}
\(\tilde{r}_-^2  \ \ind{\tau}{4}^{(-)}
+\tilde{r}_+^2 \ \ind{\tau}{4}^{(+)}
\)\\
D_1(x^\mu)&=&
\frac{\kappa\ \tilde{r}_-^2 \tilde{r}_+^2}{\pi L H^2
  \(\tilde{r}_-^3-\tilde{r}_+^3 \)}
\(\tilde{r}_+ \ \ind{\tau}{4}^{(-)}
+\tilde{r}_- \ \ind{\tau}{4}^{(+)}
\),
\ea
so that
\ba
\partial_r h^\mu_\nu=\frac{\kappa}{\pi L H^2
\(\tilde{r}_-^3-\tilde{r}_+^3 \) f(r)}\,
\left[
\frac{\tilde{r}_+^2}{r^4}\, \(\tilde{r}_-^3-r^3 \)
\ \ind{\tau}{4}^{\mu (+)}_\nu
+
\frac{\tilde{r}_-^2}{r^4}\, \(\tilde{r}_+^3-r^3 \)
\ \ind{\tau}{4}^{\mu (-)}_\nu
\right]. \label{h1'2}
\ea
In the low-energy limit, we are interested in the expression of the zero mode,
which is {\it finite} in the limit where $\epsilon \rightarrow 0$
and the same on both branes. Its contribution to the gravitational
waves is given by:
\ba
\frac{H^2}{\boxdot}\, h^{\ \mu}_{0\, \nu}(r_\pm,x^\mu)
&=&-\frac{2\tilde{\kappa}_\pm}{\boxdot}\( \ind{\tau}{4}^{\mu (\pm)}_\nu
+\frac{r_\mp^2}{r_\pm^2} \ind{\tau}{4}^{\mu (\mp)}_\nu\) \label{zero mode} \\
\tilde{\kappa}_\pm&=&
\frac{3 r_\pm^2}{2\pi L \(r_+^3-r_-^3 \)}\
\kappa.
\ea
On the brane, the Hubble parameter is fixed to $H_\pm=1/r_\pm$,
by making the choice $H=H_+$ or $H=H_-$, we can work in terms of the proper
coordinates on either of the branes. By making this choice, we have:
\ba
\tilde{\kappa}_\pm=\tilde{\kappa}=
\frac{3}{2\pi L H^2 \(r_+^3-r_-^3 \)}\
\kappa.
\ea

\subsection{Effective Newton constant}

The effective gravitational coupling constant we have derived is
exactly what one expects by naively integrated out the action
according to the usual argument
\ba
\int \d^6 x\sqrt{-g} \frac{1}{2\kappa} {}^{(6)}R &=& \int \d^4 x \d r
\d \varphi \sqrt{-g}\, \frac{1}{2\kappa} g^{\mu \nu} {}^{(4)}
R\mn+\dots \notag \\
&=&2\pi L \int \d^4 x \d r  r^2 H^2 \sqrt{-q} \frac{1}{2\kappa} \,q^{\mu \nu} {}^{(4)} R\mn +\dots \notag\\
&=&\frac{2}{3} \pi L H^2\(r_+^3-r_-^3\)
 \int \d^4 x \sqrt{-q}\, \frac{1}{2\kappa}q^{\mu \nu} {}^{(4)} R\mn \notag +\dots.
\ea
and so
\ba
\tilde{\kappa}_\pm=\tilde{\kappa}=
\frac{3}{2\pi L H^2 \(r_+^3-r_-^3 \)}\
\kappa .
\ea
This result was anticipated in ref. \cite{Mukohyama:2005yw}.
In the limit where $r_+ \gg r_-$, the zero mode couples uniquely to
the matter on the brane at $r=r_+$:
\ba
^{(0)}\!h\mupn(r_+,x^\mu)&=&
-\frac{2 \tilde{\kappa}}{\boxdot}\ \ind{\tau}{4}^{\mu
(+)}_\nu,\\
^{(0)}\!h\mupn(r_-,x^\mu)&=&
-\frac{2r_+^2}{r_-^2}
\frac{\tilde{\kappa}}{\boxdot}\ \ind{\tau}{4}^{\mu (+)}_\nu.
\ea
This is similar to the single brane limit of the
Randall-Sundrum scenario where conventional gravity is recovered on
the positive tension brane only.

\subsection{Second order}
In order to understand the behaviour of the first mode $h_1$, we need first to
constrain the remaining degree of freedom in $h_1$ by imposing the
boundary conditions on the second mode.

The first mode is given by the integration of (\ref{h1'2}). In
particular we may use the relation:
\ba
\int \frac{1}{r^m\, f(r)}\d r=\int \frac{r^{6-m}}{r^6\, f(r)} \d r
=\sum_{i/f(r_i)=0}\frac{\log \left|r-r_i\right|}{r_i^m \, f'(r_i)},
\ea
for any integer $-1\le m\le 6$, where the integral has been performed
by recognizing that the integrand is a ratio of two polynomials, with
the numerator one being of lower order.
Using this relation in the integral of (\ref{h1'2}), we hence have the
expression for the first mode
\ba
h_1&=&C_1 +\sum_{r_i} \mathcal{A}_i\, \log \left|r-r_i\right|\\
\mathcal{A}_i&=&\frac{\kappa}{ \pi L H^2 f'(r_i)\(\rrm^3-\rrp^3\)}
\left[
\frac{\rrp^2}{r_i^4}\(\rrm^3-r_i^3\) {}^{(4)\!}\tau^{(+)}
+\frac{\rrm^2}{r_i^4}\(\rrp^3-r_i^3\) {}^{(4)\!}\tau^{(-)}
\right].
\ea
We may now use this expression to derive the second mode, which using
the equation (\ref{modes eq}), is of the form:
\ba
f \partial_r h_2=g_2&=&\frac{1}{r^4}\left[
D_2-\int r^2 h_1 \d r
\right]\notag \\
&=&\frac{1}{r^4}\left[
D_2-\frac 1 3 C_1 r^3+z(r)
\right]\label{g2}\\
z(r)&=&-\frac 1 3 \sum_{r_i}\mathcal{A}_i
\left(
-r\(\frac 1 3 r^2+ \frac 1 2 r_i r+r_i^2\)
+\(r^3-r_i^3\)\log \left|r-r_i\right|
\right).
\ea
From the boundary conditions, $\partial_r h_2$ should vanish on both branes at
$\rt$. This constraint fixes the constants $C_1$ and $D_2$ to
\ba
C_1&=&\frac{3\(z(\rrm)-z(\rrp)\)}{\rrm^3-\rrp^3}\label{C1}\\
D_2&=&\frac{\rrp^3 z(\rrm)-\rrm^3 z(\rrp)}{\rrm^3-\rrp^3},\label{D2}
\ea
leading to the following expression for the first mode:
\ba
h_1=\frac{3\(z(\rrm)-z(\rrp)\)}{\rrm^3-\rrp^3}
+\sum_{r_i}\mathcal{A}_i \log \left|r-r_i\right|.
\ea
We may check that in general, the first mode is not finite in the
thin brane limit $\epsilon \rightarrow 0$. In
particular, the dominant contribution to the first mode on the branes
is logarithmically divergent
\ba
h_1(r_\pm)&=&\mathcal{A}_\pm \log \epsilon
+\mathcal{O}(\epsilon^0)\notag \\
&=&-\frac{\kappa \log \epsilon}{ \pi L H^2 r_\pm^2 \left|f'(r_\pm)\right|} \
{}^{(4)\!}\tau^{(\pm)}
+\mathcal{O}(\epsilon^0). \label{first mode}
\ea
Putting all this together, we have the following effective four dimensional equations of motion on each brane
\be
h_{\pm}=-\frac{2\tilde{\kappa}}{\boxdot}\  \( \ind{\tau}{4}^{
(\pm)} +\frac{r_{\mp}^2}{r_{\pm}^2}\ind{\tau}{4}^{
(\mp)} \) - \frac{\kappa \log \epsilon}{ \pi L H^2 r_\pm^2 \left|f'(r_\pm)\right|} \
{}^{(4)\!}\tau^{(\pm)}
+\mathcal{O}(\epsilon^0) .
\ee

\subsection{Renormalization and EFT}

The presence of the logarithmic dependence on the cutoff suggests an
inherent model dependence in the form of the solutions. That is,
it will be necessary to specify certain features of the brane physics
in order to determine a unique solution in the bulk. At first sight
one may not be so concerned about this, since the dependence is only
logarithmic in the cutoff. However, this is a feature of the fact that
we are doing linearized perturbations. The logarithmic divergence of
the metric perturbations near the brane is in fact a signal of the
onset of a generic anisotropic Kasner-like singularity which the
conical singularity is unstable to. Consequently we only expect this
dependence on $\epsilon$ to get worse at higher orders. We stress that this logarithmic divergence is not the same as the more familiar infrared logarithmic divergence of massless scalar fields on two dimensional spacetimes, in fact we anticipate that for any codimension we will still find a logarithmic dependence for gravitational waves, where the infrared behaviour falls of as a power for higher codimension. 

In EFT we are used to the idea that if we are interested in physics at
a given energy scale, we can integrate out the modes whose masses are
much larger than this energy scale, and the net effect is to just
renormalize various counterterms in the effective action. It is
tempting to apply the same philosophy here, the cutoff $\epsilon$ is
associated with an energy scale $1/\epsilon$ at which modes which
describe the brane itself will become excited. If we concentrate on
physics at energy scales well below this, for instance at scales set
by the size of the extra dimensions for which the higher derivative
terms we have been discussing are still important, then it is natural
to expect that all dependence on the scale $1/\epsilon$ can be
renormalized or canceled by various counterterms localized either on
the brane or in the bulk.

Consider
first the possibility that the $\log \epsilon$ may be absorbed in
brane counterterms that are local functions of the metric, curvature
invariants and the matter degrees of freedom on the brane. This
minimal possibility seems natural given that the high energy brane
physics is localized at the brane itself. 
Such terms would correspond to redefinitions of
the stress energy on each brane. The problem is that the metrics on
each brane depend on the same combination of ${}^{(4)}\tau^{(\pm)}$ at
leading order and so any redefinition would give the same contribution
on each brane. However in general the log divergence on each brane is
different and so no simple renormalization of the stress energy on
each brane will cancel the logarithmic dependence.  

The next possibility is to include in the brane action functions of
the extrinsic curvature. This is only consistent as a boundary
condition if we also include higher order derivative terms in the bulk
that increase the order of the Cauchy problem. Adding counterterms in
the
bulk as well as on the brane can allow one to cancel the $\log \epsilon$
divergence on each brane, but closer inspection shows that this is
typically always at the price of reintroducing it in the bulk. The key
point is
that as long as the fact that the perturbations diverge logarithmically near the
brane is unaffected by the bulk counterterms, the logarithmic divergences will still show up at the brane.  
However, we have not as yet performed an exhaustive analysis of all the possible counterterms that could be used. In section \ref{GB} we discuss an alternative approach based on using
GB terms and show that also does not remove the $\log \epsilon$
divergences.

\subsubsection*{Position of the brane}

This apparent difficulty at reconciling these results with the
intuition from effective field theory may be a consequence of the fact
that the boundary condition approach we have used in section
\ref{boundaryconditions2} is not adequate. For instance, in this
approach we are interpreting the metric evaluate at
$r=\tilde{r}_{\pm}$ to be the metric of the brane. However in practice
the brane is a thick object smoothed over a region of width
$\epsilon$. We have made the assumption that the variation of the
metric across the brane thickness is suppressed by $\epsilon$. In fact
this is not necessarily the case, suppose for example we assume that
on the region $|r-r_{\pm}|<\epsilon$, the metric perturbation
varies as $h \approx A+B (r-r_{\pm})^2$. Matching this form to the
known form for $|r-r_{\pm}|>\epsilon$ shows us that
$h(r=\tilde{r}_{\pm})-h(r=r_{\pm}) \approx \frac{\kappa}{LH^2
  r_{\pm}^2 |f'(r_{\pm})|}{}^{(4)}\tau^{(\pm)}$. This variation is
precisely of the same order as the first correction. The implication
is that the coefficient of this correction will vary depending on
where within the brane we choose to call the `brane
position'. In some sense one can renormalize the $\log \epsilon$
dependence into the `position' of the brane. Allowing for more
complicated evolution of the metric in the brane regime will allow for
more possibilities to absorb the $\log$ dependence. However, we are
left with the same conclusion, we need to
specify in more detail the physics of the brane in the region
$|r-r_{\pm}|<\epsilon$ in order to make predictions for even the
leading order six dimensional corrections to the gravitational wave
propagation. 

What seems to be lacking is a consistent way of separating the physics
corresponding to different scales in the manner of effective field
theory. It seems clear that something along these lines needs to be
developed before a good understanding of the dynamics of codimension
two branes can be achieved. Interesting work 
in this direction has
been done in \cite{Goldberger:2004jt}, and extending 
these ideas
to arbitrary codimension branes seems to be crucial to capturing
the essential physics of higher codimension braneworlds
\cite{TolleyBurgess}.

\subsection{Higher orders in the derivative expansion}
\label{sec higher order}

This derivative expansion may be continued to arbitrary high order.
It is important to check that no further divergences are introduced
so that our approximations are self consistent. Actually we can show
that the next orders are regular in the limit
$\epsilon \rightarrow 0$. Using the expression
(\ref{g2}) with the relations (\ref{C1}) and (\ref{D2}) for the
integration constants, we see that $\partial_r h_2$ is regular
everywhere.  Since $\partial_r h_2$ vanishes on the branes,
$h_2$ must be regular at that point as well, and hence the second mode
is regular everywhere. Furthermore, if for a given mode $n\ge 2$,
$h_n$ is regular everywhere, then
$f (r) \partial_r h_{n+1}(r)=g_{n+1}(r)
=\frac{1}{r^4}\int_r^{r\star} r' h_n(r')\d r'$
is similarly finite everywhere. Since $\partial_r h_{n+1}$ vanishes
on the branes, the next mode $h_{n+1}=\int g/f \d r$
is therefore finite on the brane as well and regular
everywhere. Therefore only the first mode has a logarithmic
divergence, all further modes are finite in the $\epsilon \rightarrow 0$ limit on the branes and
this derivative expansion is hence well-defined and may be continued
to higher orders.

\subsection{Compactification with one brane}

It is possible to take a one-brane limit whereby the background
constants are chosen so that the tension of one of the branes is zero
and at the perturbed level the associated stress energy vanishes. We
find that this is a well defined limit since the metric perturbations
do not diverge at the smooth pole.

\section{Kaluza-Klein limit}
\label{sectionV}
\subsection{Limit of the General solution}
The general solutions we have considered so far which describe
warped compactifications, contain the more familiar Kaluza-Klein
compactifications as a special limit. In particular we may obtain
solutions describing $\d S^4 \times S^2$ as follows: Redefine
$r=H^{-1}+\sigma \rho$ and specify the constants $\mu$, $b$ and
$L$ as
\be
\hspace{-10pt}
\mu = \frac{1}{15}\Lambda H^{-5}-\frac{1}{6}b^2H^3, \hspace{20pt}
\frac{1}{12}b^2H^6 =\frac{1}{6} \Lambda H^{-2}+\sigma^2-1, \hspace{20pt}
L = \frac{R_c}{\sigma},
\ee
where
\be
\frac{1}{R_c^2}=2\Lambda-9H^2.
\ee
Then on taking the limit $\sigma \rightarrow 0$ we find the metric
\be
\d s^2=\d s^2_{\d S^4}+\tilde{f}(\rho)^{-1}\d\rho^2
+R_c^2\, \tilde{f}(\rho) \d\varphi^2,
\ee
with $\tilde{f}(\rho)=f(\rho)/\sigma^2=1-\rho^2/R_c^2$.
This describes a direct product of four-dimensional de Sitter with Hubble
constant $H$ and a two-sphere with curvature radius $R_c$. After a
suitable gauge transformation the gauge field is given by
$A=-(bH^4R_c) \rho \d \varphi$ corresponding to a constant flux.
Solutions describing $Minkowski^4 \times S^2$ and $A\d S^4 \times S^2$
are similarly obtained by taking $H^2=0$ and $H^2<0$ respectively.

\subsection{Kaluza-Klein solution}

In the special KK limit, the exact behavior of the different modes may be
computed exactly. This provides a useful test on the procedure we
have presented, but allows us as well to understand the behavior
and the nature of the divergence beyond the zeroth order.
The evolution equation in this limit $H r \rightarrow 1$ is simply
\ba
 \partial_\rho \(\tilde{f} \partial_\rho h\)&=&-\Box h \\
\left. \partial_\rho h\right|_{\pm R_c\mp \epsilon}
&=&\pm \frac{\kappa}{ \pi R_c \, \tilde{f}(\pm R_c\mp
\epsilon)}\ind{\tau}{4}^{(\pm)},\label{bdyKK}
\ea
where the four-dimensional indices $\mu,\nu$ have been omitted to
simplify the notation. The solution is therefore of the form:
\ba
h=C_1 \mathcal{P}_m (\rho/R_c)+ C_2 \mathcal{Q}_m (\rho/R_c),
\label{sol KK}
\ea
where $\mathcal{P}_m$ and $\mathcal{Q}_m$ are the Legendre polynomials and
$m=-1/2 +\sqrt{1/4+R_c^2 \ \Box}$. The two integration functions
$C_{1,2}$ can be fixed using the boundary condition
(\ref{bdyKK}). Using these, the gravitational waves
on each brane $h^{\pm}$ are sourced by the matter perturbations in
the following way:
\ba
h^{\pm}=-\frac{\kappa}{2\pi}
\left[
\(
{\log \frac{\epsilon}{2 R_c}}
\color{black}{+}H_m+H_{-m-1}\)\ind{\tau}{4}^{(\pm)}
+
\frac{\pi}{\sin m \pi } \, \ind{\tau}{4}^{(\mp)}
 \right],\label{solKK2}
\ea
where $H_m$ is the $m^{\text{th}}$ Harmonic number: $H_m=\sum_{k=1}^m
k^{-1}$ which for noninteger $m$ can be defined by $H_m=\int_0^1
\frac{1-x^m}{1-x} \ \d x$.
To leading order, we therefore have:
\ba
h^{\pm}&=&
-\frac{\kappa}{2\pi\, R_c^2
  \Box}\left[\ind{\tau}{4}^{(+)}+\ind{\tau}{4}^{(-)} \right]
-\frac{\kappa}{2\pi}\left[\(1+\log \frac{\epsilon}{2
    R_c}\)\ind{\tau}{4}^{(\pm)}+\ind{\tau}{4}^{(\mp)}\right]\label{solKK3}\\
&& +\frac{\kappa R_c^2 \Box}{2\pi}
\left[\ind{\tau}{4}^{(\pm)}+\(1-\frac{\pi^2}{6}\)\ind{\tau}{4}^{(\mp)}\right]
+\cdots\notag
\ea

The essential features are the same as those observed
in the general case. First we may emphasize that only the first mode
diverges and its divergence is very mild since it is only logarithmic
in the cutoff parameter $\epsilon$. Another important feature is that
this divergence only couples to the matter on the specific brane and
{\it not} to the matter content of the other brane.

This represents a consistency check on the procedure used in the
low-energy limit to derive each modes separately, since we recover
the same result. We recover for both the leading and first order in the expansion, the results obtained in
(\ref{zero mode}) and (\ref{first mode}) corresponds precisely with
the result in (\ref{solKK3}) in the limit $\sigma \rightarrow 0$.

\section{Effect of Gauss-Bonnet terms}

\label{GB}

Faced with the fact that higher codimension distributional sources are
typically singular in GR, one popular approach to dealing with this is
to include higher derivative terms in the bulk which allow
distributional sources to be consistent with the equations of
motion. In particular, for codimension two branes the main focus has
been on Gauss-Bonnet (GB) terms. The effect of introducing these terms in
the bulk is to modify the boundary conditions in such a way that the
singular solutions may be discarded \cite{Davis:2002gn}. For instance a simple analysis of
a codimension two brane in uncompactified Minkowski space shows that
at the perturbed level the bulk equations of motion are unaffected and
hence the logarithmically diverging solution is still present, however
the boundary conditions are modified so that we may ignore this
solution. In this section we point out that as soon as we consider
codimension two branes in a compactified space this procedure
fails. The reason is simply that if we restrict ourself to the
solution which is regular at one brane, this solution will inevitably
diverge at the other brane (or pole if no brane is present).

Introducing a GB term in the six dimensional action:
\ba
S_{\text{GB}}=\frac{\alpha}{2\sqrt{\kappa}}\int \d ^6 x\sqrt{-g} \left[
^{(6)}R^{ABCD}\, {}^{(6)\!}R_{ABCD}-4\,\,^{(6)\!\!}R^{AB}\, {}^{(6)\!}R_{AB}+{}^{(6)\!}R^2
\right],
\ea
where $\alpha$ is a dimensionless parameter, the presence of this term
modifies both the bulk Einstein equation and the junction condition. In
particular, the modified Einstein equation is:
\ba
^{(6)\!}G_{AB}+\alpha \sqrt{\kappa} \(2\, \mathcal{R}^{\text{GB}}_{AB}-\frac 1 2
\mathcal{R}^{\text{GB}}\, g_{AB}\)=
-\Lambda \, g_{AB}+\(
F_A^{\ \ C}\, F_{BC}- \frac 1 4 F^{CD}F_{CD}\, g_{AB}
\), \label{eq with alpha}
\ea
with
\ba
\mathcal{R}^{\text{GB}}_{AB}=
\,{}^{(6)\!\!}R\,{}^{(6)\!} R_{AB}-2 \,^{(6)\!}R_{A C}\, ^{(6)\!}R^C_{\ \
B}-2 \, ^{(6)\!}R^{CD}\, ^{(6)\!}R_{ACBD}+\, ^{(6)\!}R_A^{\ \ DEF}\, ^{(6)\!}R_{B DEF}.
\ea
In the Kaluza-Klein limit, we consider the bulk metric of the form
\be
\d s^2=\d s^2_{\d S^4}+\tilde{f}(\rho)^{-1}\d\rho^2
+L^2\, \tilde{f}(\rho) \d\varphi^2. \label{metric KK limit}
\ee
The limit $L = R_c$ corresponds to the previous situation where the
brane tension vanishes, and the background geometry is smooth without
any singularity. Taking $L<R_c$ corresponds to the rugby ball geometry
corresponding to two equal tension branes at each pole.
The contribution of the GB term in this geometry
is:
\ba
 \(2\, \mathcal{R}^{\text{GB}}{}^A_{\ B}-\frac 1 2
\mathcal{R}^{\text{GB}}\, \delta^A_{\ B}\)=
\(
\begin{array}{cc}
-\frac{12 H^2}{R_c^2}\, \delta^\mu_{\ \nu} & \\
& -12 H^4 \delta^x_{\ y}
\end{array}
\),
\ea
where the indices $x,y$ run over the two extra dimensions:
$x,y=\varphi, \rho$. These terms may hence be
interpreted as a redefinition of the background parameters, and in
particular, the metric (\ref{metric KK limit}) is a solution of the
modified Einstein equation (\ref{eq with alpha}) if
\ba
\frac{1}{R_c^2}&=&\frac{1}{1+12 \alpha \sqrt{\kappa} H^2}\left[
2 \Lambda -\(9+12 \alpha \sqrt{\kappa} H^2\)H^2
\right]\\
A_B\, \d x^B&=&-\sqrt{1-3 R_c^2H^2+12 \alpha \sqrt{\kappa} H^2
  \(R_c^{\, -2}-H^2\)}\ \rho\,  \d \varphi.
\ea
In this limit, the form of the background solution is unaffected by the
presence of the GB term, and in particular in the solution that
describes $Minkowski^4 \times S^2$, ie. when $H^2=0$, the GB terms
vanish at this order.
If we concentrate on this specific solution, at the perturbed level,
the contribution of these terms is simply:
\ba
\delta \mathcal{R}^{\text{GB}}{}^{\, \mu}_{\ \,
  \nu}=-\frac{1}{R_c^2}\, \Box h^\mu_{\
\nu},
\ea
and all other component vanish.
The equation of motion of the gravitation waves
is only very slightly modified:
\ba
\partial_\rho \(\tilde{f} \, \partial_\rho \,
h^\mu_{\ \nu}\)=-\(1-2\frac{\alpha \sqrt{\kappa}}{R_c^2}\) \Box\, h^\mu_{\ \nu}.
\ea
The gravitational waves hence behave identically as in (\ref{sol KK}),
with only a modification of the parameter $m$:
$m=-\frac 1 2 +\sqrt{\frac 1 4+R_c^2({1-2\alpha \sqrt{\kappa}/R_c^2}) \
  \Box}$. Although the boundary conditions might be modified we argue
that the addition of this term can not remove the logarithmic
divergence obtained in (\ref{solKK2}, \ref{solKK3}). The key point is
that the form of the solution (\ref{sol KK}) remains completely
unaffected by the addition of the GB term. Both Legendre polynomials
have a logarithmic divergence when $\rho \rightarrow \pm R_c$ and
the solution is therefore of the form:
\ba
h(\rho=R_c-\epsilon)&\rightarrow& -\frac{C_2}{2}\ \log \frac{\epsilon}{2
  R_c}\\
h(\rho=-R_c+\epsilon)&\rightarrow& \(\frac{C_1\, \sin m
  \pi}{\pi}+\frac{C_2\, \cos m \pi}{2}\)\ \log \frac{\epsilon}{2 R_c}.
\ea
Requiring that the logarithmic divergence cancels on both branes,
would fix both parameters $C_1$ and $C_2$ to zero, giving
rise to the trivial solution. This is clearly an unphysical
restriction, and so we conclude that the GB terms do not regulate the
branes.

Although we have reached this conclusion for the case of linear
perturbations around the Kaluza-Klein solution, it should be clear
that the same result will occur in general. In fact it is a
fundamental feature of the GB terms that they do not change the nature
of the Cauchy problem (i.e. the differential equations remain second
order) \cite{Deruelle:2003ck}, as a result we are always dealing with `two solutions' both of
which will diverge at one or both branes even in the nonlinear
case. Furthermore the GB terms will not significantly alter the
asymptotic form of the solutions near the brane since they only become
significant when the curvature becomes large, and for conical
singularities the curvature remains finite and arbitrarily small up to
the singularity itself. Demanding that the metric is conical at both
branes is too restrictive a condition on the space of solutions that
we will be left with essentially a trivial solution. This fact casts
serious doubt on whether it makes sense to regulate codimension two
branes with GB terms even in the uncompactified case, since if the
regulating physics is local then this case should be no different that
the compactified case in the limit in which the extra dimensions are
very large.

\section{Conclusion}

\label{conclusion}

In the context of an explicit six dimensional braneworld model, we have considered
the effective equations describing the propagation of bulk
gravitational waves induced by matter sources living on the branes.
We have shown that one obtains ordinary gravity at low energies,
justifying the boundary condition approach we have used, and have
explicitly determined the leading order modifications due to six
dimensional effects. We have seen that there is an apparent model
dependence in these corrections and have discussed the possibility of
renormalizing this in brane or bulk counterterms, in accordance with
expectations from effective field theory. Although it seems plausible
that the cutoff dependence may be absorbed into counterterms localized
near the brane, we find that it is technically difficult to do so. 
Our work highlights the need to develop a more consistent picture of how to infer bulk dynamics from the boundary conditions imposed by the brane physics using the ideas of effective field theory. We
have shown that the use
of Gauss-Bonnet terms to regulate codimension two branes is
inconsistent when the extra dimensions are compact, suggesting in
fact that even in the uncompactified case they represent an
unphysical regularization. Since we focussed on tensor perturbations,
much of the analysis was straightforward, but many questions remain
such as how to deal with the boundary conditions in the scalar
sector, and what happens to the logarithmic divergences in the
nonlinear theory. Many of the features of codimension two branes
discussed here will be present for higher codimension branes, although
the codimension two is a specific case that can not be treated the same
way as higher codimensional branes \cite{Kakushadze:2001bd}.

\section{Acknowledgements}
We would like to thank I.~Antoniadis, R.~Brandenberger, C.~P.~Burgess, J.~Cline,
O.~Corradini, I.~Navarro and G.~Tasinato for useful discussions and comments. CdR
is funded by a grant from the Swiss National Science Foundation. AJT is supported in part by
US Department of Energy Grant DE-FG02-91ER40671.

\bibliographystyle{is-unsrt}

\begin{thebibliography}{is-unsrt}

\bibitem{Mukohyama:2005yw}
  S.~Mukohyama, Y.~Sendouda, H.~Yoshiguchi and S.~Kinoshita,
  ``Warped flux compactification and brane gravity,''
  JCAP {\bf 0507}, 013 (2005)
  [arXiv:hep-th/0506050].

\bibitem{Arkani-Hamed:1998rs}
  N.~Arkani-Hamed, S.~Dimopoulos and G.~R.~Dvali,
  ``The hierarchy problem and new dimensions at a millimeter,''
  Phys.\ Lett.\ B {\bf 429}, 263 (1998)
  [arXiv:hep-ph/9803315].

\bibitem{Antoniadis:1998ig}
  I.~Antoniadis, N.~Arkani-Hamed, S.~Dimopoulos and G.~R.~Dvali,
  ``New dimensions at a millimeter to a Fermi and superstrings at a TeV,''
  Phys.\ Lett.\ B {\bf 436}, 257 (1998)
  [arXiv:hep-ph/9804398].
  
\bibitem{Goldberger:2004jt}
  W.~D.~Goldberger and I.~Z.~Rothstein,
  ``An effective field theory of gravity for extended objects,''
  arXiv:hep-th/0409156.
  
  W.~D.~Goldberger and I.~Z.~Rothstein,
  ``Dissipative Effects in the Worldline Approach to Black Hole Dynamics,''
  arXiv:hep-th/0511133.


  



 \bibitem{Aghababaie:2003wz}
 Y.~Aghababaie, C.~P.~Burgess, S.~L.~Parameswaran and F.~Quevedo,
 ``Towards a naturally small cosmological constant from branes in 6D
 supergravity,''
 Nucl.\ Phys.\ B {\bf 680}, 389 (2004)
 [arXiv:hep-th/0304256];



\bibitem{Rubakov:1983bz}
 V.~A.~Rubakov and M.~E.~Shaposhnikov,
 ``Extra Space-Time Dimensions: Towards A Solution To The Cosmological
 Constant Problem,''
 Phys.\ Lett.\ B {\bf 125}, 139 (1983);
 %
  G.~Dvali, G.~Gabadadze and M.~Shifman,
  ``Diluting cosmological constant in infinite volume extra dimensions,''
  Phys.\ Rev.\ D {\bf 67}, 044020 (2003)
  [arXiv:hep-th/0202174];
%
  O.~Corradini, A.~Iglesias and Z.~Kakushadze,
  ``Toward solving the cosmological constant problem?,''
  Int.\ J.\ Mod.\ Phys.\ A {\bf 18}, 3221 (2003)
  [arXiv:hep-th/0212101];
%
S.~M.~Carroll and M.~M.~Guica,
``Sidestepping the cosmological constant with football-shaped extra
dimensions,''
arXiv:hep-th/0302067;
%
%
 I.~Navarro,
 ``Codimension two compactifications and the cosmological constant  problem,''
 JCAP {\bf 0309}, 004 (2003)
 [arXiv:hep-th/0302129];
%
%
 I.~Navarro,
 ``Spheres, deficit angles and the cosmological constant,''
 Class.\ Quant.\ Grav.\  {\bf 20}, 3603 (2003)
 [arXiv:hep-th/0305014];
%
  O.~Corradini, A.~Iglesias and Z.~Kakushadze,
  ``Diluting solutions of the cosmological constant problem,''
  Mod.\ Phys.\ Lett.\ A {\bf 18}, 1343 (2003)
  [arXiv:hep-th/0305164];
%
  H.~P.~Nilles, A.~Papazoglou and G.~Tasinato,
  ``Selftuning and its footprints,''
  Nucl.\ Phys.\ B {\bf 677}, 405 (2004)
  [arXiv:hep-th/0309042];
%
 J.~Vinet and J.~M.~Cline,
 ``Can codimension-two branes solve the cosmological constant problem?,''
 Phys.\ Rev.\ D {\bf 70}, 083514 (2004)
 [arXiv:hep-th/0406141];
%
 J.~Garriga and M.~Porrati,
 ``Football shaped extra dimensions and the absence of self-tuning,''
 JHEP {\bf 0408}, 028 (2004)
 [arXiv:hep-th/0406158];
%
 J.~Vinet and J.~M.~Cline,
 ``Codimension-two branes in six-dimensional supergravity and the cosmological
 constant problem,''
 Phys.\ Rev.\ D {\bf 71}, 064011 (2005)
 [arXiv:hep-th/0501098];
%
 J.~M.~Schwindt and C.~Wetterich,
 ``The cosmological constant problem in codimension-two brane models,''
 Phys.\ Lett.\ B {\bf 628}, 189 (2005)
 [arXiv:hep-th/0508065].
%

 \bibitem{Burgess:2005wu}
 C.~P.~Burgess,
 ``Supersymmetric large extra dimensions and the cosmological constant
 problem,''
 arXiv:hep-th/0510123.




\bibitem{Geroch:1987qn}
  R.~Geroch and J.~H.~Traschen,
  ``Strings And Other Distributional Sources In General Relativity,''
  Phys.\ Rev.\ D {\bf 36}, 1017 (1987).



\bibitem{Corradini:2002ta}
  O.~Corradini, A.~Iglesias, Z.~Kakushadze and P.~Langfelder,
  ``A remark on smoothing out higher codimension branes,''
  Mod.\ Phys.\ Lett.\ A {\bf 17}, 795 (2002)
  [arXiv:hep-th/0201201].






\bibitem{Vilenkin:1981zs}
 A.~Vilenkin,
 ``Gravitational Field Of Vacuum Domain Walls And Strings,''
 Phys.\ Rev.\ D {\bf 23} (1981) 852;
 A.~Vilenkin and E.~P.~S.~Shellard, Cosmic Strings and Other Topological Defects, Cambridge University Press.




\bibitem{Gherghetta:2000qi}
 T.~Gherghetta and M.~E.~Shaposhnikov,
 ``Localizing gravity on a string-like defect in six dimensions,''
 Phys.\ Rev.\ Lett.\  {\bf 85}, 240 (2000)
 [arXiv:hep-th/0004014].




\bibitem{Bazeia:2003qt}
 D.~Bazeia, J.~Menezes and R.~Menezes,
 ``New global defect structures,''
 Phys.\ Rev.\ Lett.\  {\bf 91}, 241601 (2003)
 [arXiv:hep-th/0305234].




\bibitem{Bostock:2003cv}
 P.~Bostock, R.~Gregory, I.~Navarro and J.~Santiago,
 ``Einstein gravity on the codimension 2 brane?,''
 Phys.\ Rev.\ Lett.\  {\bf 92}, 221601 (2004)
 [arXiv:hep-th/0311074].

\bibitem{Corradini:2001su}
  O.~Corradini and Z.~Kakushadze,
  ``A solitonic 3-brane in 6D bulk,''
  Phys.\ Lett.\ B {\bf 506}, 167 (2001)
  [arXiv:hep-th/0103031];
%
  O.~Corradini, A.~Iglesias, Z.~Kakushadze and P.~Langfelder,
  ``Gravity on a 3-brane in 6D bulk,''
  Phys.\ Lett.\ B {\bf 521}, 96 (2001)
  [arXiv:hep-th/0108055];
%
 S.~Kanno and J.~Soda,
 ``Quasi-thick codimension 2 braneworld,''
 JCAP {\bf 0407}, 002 (2004)
 [arXiv:hep-th/0404207];
%
 P.~Wang and X.~H.~Meng,
 ``Codimension two branes in Einstein-Gauss-Bonnet gravity,''
 Phys.\ Rev.\ D {\bf 71}, 024023 (2005)
 [arXiv:hep-th/0406170];
 %
  S.~Kanno and J.~Soda,
  ``On the higher codimension braneworld,''
  arXiv:gr-qc/0410067;
%
 G.~Kofinas,
 ``Conservation equation on braneworlds in six dimensions,''
 Class.\ Quant.\ Grav.\  {\bf 22}, L47 (2005)
 [arXiv:hep-th/0412299];
%
 E.~Papantonopoulos and A.~Papazoglou,
 ``Brane-bulk matter relation for a purely conical codimension-2 brane
 world,''
 JCAP {\bf 0507}, 004 (2005)
 [arXiv:hep-th/0501112];
%
%
 C.~Charmousis and R.~Zegers,
 ``Matching conditions for a brane of arbitrary codimension,''
 JHEP {\bf 0508}, 075 (2005)
 [arXiv:hep-th/0502170].





\bibitem{Kofinas:2005py}
 G.~Kofinas,
 ``On braneworld cosmologies from six dimensions, and absence thereof,''
 arXiv:hep-th/0506035.





\bibitem{Arkani-Hamed:1999hk}
 N.~Arkani-Hamed, S.~Dimopoulos, G.~R.~Dvali and N.~Kaloper,
 ``Infinitely large new dimensions,''
 Phys.\ Rev.\ Lett.\  {\bf 84}, 586 (2000)
 [arXiv:hep-th/9907209];
%
 C.~Csaki and Y.~Shirman,
 ``Brane junctions in the Randall-Sundrum scenario,''
 Phys.\ Rev.\ D {\bf 61}, 024008 (2000)
 [arXiv:hep-th/9908186];
%
%
 A.~E.~Nelson,
 ``A new angle on intersecting branes in infinite extra dimensions,''
 Phys.\ Rev.\ D {\bf 63}, 087503 (2001)
 [arXiv:hep-th/9909001];
 %
  J.~E.~Kim, B.~Kyae and H.~M.~Lee,
  ``Localized gravity and mass hierarchy in D = 6 with Gauss-Bonnet term,''
  Phys.\ Rev.\ D {\bf 64}, 065011 (2001)
  [arXiv:hep-th/0104150];
%
%
 E.~Gravanis and S.~Willison,
 ``Intersecting hyper-surfaces in dimensionally continued topological  density
 gravitation,''
 J.\ Math.\ Phys.\  {\bf 45}, 4223 (2004)
 [arXiv:hep-th/0306220];
%
 E.~Gravanis and S.~Willison,
 ``Intersecting hypersurfaces, topological densities and Lovelock gravity,''
 arXiv:gr-qc/0401062;
%
  H.~M.~Lee and G.~Tasinato,
  ``Cosmology of intersecting brane world models in Gauss-Bonnet gravity,''
  JCAP {\bf 0404} (2004) 009
  [arXiv:hep-th/0401221];
%
 I.~Navarro and J.~Santiago,
 ``Higher codimension braneworlds from intersecting branes,''
 JHEP {\bf 0404}, 062 (2004)
 [arXiv:hep-th/0402204];
%
 N.~Kaloper,
 ``Origami world,''
 JHEP {\bf 0405}, 061 (2004)
 [arXiv:hep-th/0403208];
%
 E.~Gravanis and S.~Willison,
 ``Intersecting membranes in AdS and Lovelock gravity,''
 arXiv:hep-th/0412273.




\bibitem{Zwiebach:1985uq}
  B.~Zwiebach,
  ``Curvature Squared Terms And String Theories,''
  Phys.\ Lett.\ B {\bf 156}, 315 (1985);
%
  D.~J.~Gross and J.~H.~Sloan,
  Nucl.\ Phys.\ B {\bf 291}, 41 (1987).



\bibitem{Navarro:2004di}
  I.~Navarro and J.~Santiago,
  ``Gravity on codimension 2 brane worlds,''
  JHEP {\bf 0502}, 007 (2005)
  [arXiv:hep-th/0411250].




\bibitem{Gregory:1995qh}
 R.~Gregory,
 ``Cosmic p-Branes,''
 Nucl.\ Phys.\ B {\bf 467}, 159 (1996)
 [arXiv:hep-th/9510202].




\bibitem{Kanti:2001vb}
  P.~Kanti, R.~Madden and K.~A.~Olive,
  ``A 6-D brane world model,''
  Phys.\ Rev.\ D {\bf 64}, 044021 (2001)
  [arXiv:hep-th/0104177].

\bibitem{Gogberashvili:2003ys}
 M.~Gogberashvili and D.~Singleton,
 ``Brane in 6D with increasing gravitational trapping potential,''
 Phys.\ Rev.\ D {\bf 69}, 026004 (2004)
 [arXiv:hep-th/0305241];
%
 P.~Midodashvili,
 ``Brane in 6D and localization of matter fields,''
 arXiv:hep-th/0308003;
%
 P.~Midodashvili and L.~Midodashvili,
 ``Gravitational localization of matters in 6D,''
 Europhys.\ Lett.\  {\bf 65}, 640 (2004)
 [arXiv:hep-th/0308039];
%
 P.~Midodashvili,
 ``Gravitational trapping of the bulk fields in 6D,''
 arXiv:hep-th/0308051;
%
  Y.~Aghababaie {\it et al.},
  ``Warped brane worlds in six dimensional supergravity,''
  JHEP {\bf 0309}, 037 (2003)
  [arXiv:hep-th/0308064].




\bibitem{Hayakawa:2003qm}
  S.~Hayakawa, D.~Ida, T.~Shiromizu and T.~Tanaka,
  ``Gravitation in the codimension two brane world,''
  Prog.\ Theor.\ Phys.\ Suppl.\  {\bf 148}, 128 (2003);
%
%
\bibitem{Cline:2003ak}
 J.~M.~Cline, J.~Descheneau, M.~Giovannini and J.~Vinet,
 ``Cosmology of codimension-two braneworlds,''
 JHEP {\bf 0306}, 048 (2003)
 [arXiv:hep-th/0304147];
%
  J.~Vinet,
  ``Generalised cosmology of codimension-two braneworlds,''
  Int.\ J.\ Mod.\ Phys.\ A {\bf 19}, 5295 (2004)
  [arXiv:hep-th/0408082];
%
 C.~Charmousis and R.~Zegers,
 ``Einstein gravity on an even codimension brane,''
 Phys.\ Rev.\ D {\bf 72}, 064005 (2005)
 [arXiv:hep-th/0502171];
%
 E.~Papantonopoulos and A.~Papazoglou,
 ``Cosmological evolution of a purely conical codimension-2 brane world,''
 JHEP {\bf 0509}, 012 (2005)
 [arXiv:hep-th/0507278].



\bibitem{Garriga:1999yh}
  J.~Garriga and T.~Tanaka,
  ``Gravity in the brane-world,''
  Phys.\ Rev.\ Lett.\  {\bf 84}, 2778 (2000)
  [arXiv:hep-th/9911055].


 \bibitem{TolleyBurgess}
  C.~P.~Burgess and A.~J.~Tolley, work in progress.


\bibitem{Davis:2002gn}
 S.~C.~Davis,
 ``Generalised Israel junction conditions for a Gauss-Bonnet brane world,''
 Phys.\ Rev.\ D {\bf 67}, 024030 (2003)
 [arXiv:hep-th/0208205];
%
 E.~Gravanis and S.~Willison,
 ``Israel conditions for the Gauss-Bonnet theory and the Friedmann equation on
 the brane universe,''
 Phys.\ Lett.\ B {\bf 562}, 118 (2003)
 [arXiv:hep-th/0209076].






\bibitem{Deruelle:2003ck}
 N.~Deruelle and J.~Madore,
 ``On the quasi-linearity of the Einstein- 'Gauss-Bonnet' gravity field
 equations,''
 arXiv:gr-qc/0305004.

\bibitem{Kakushadze:2001bd}
  Z.~Kakushadze,
  ``Orientiworld,''
  JHEP {\bf 0110}, 031 (2001)
  [arXiv:hep-th/0109054].
  





























\end{thebibliography}

\end{document}